\documentclass[aps,twocolumn,groupedaddress,showpacs,showkeys]{revtex4}

\usepackage{makeidx}
\usepackage{amsmath}
\usepackage{amsthm}
\usepackage{amssymb}
\usepackage{amsxtra}
\usepackage{amscd}
\usepackage[mathscr]{eucal}
\usepackage{boxedminipage}
\usepackage{bbm}
\usepackage{epsfig}
\usepackage{epic,eepic}
\usepackage{latexsym}
\usepackage{graphics}

\newcommand{\ket}[1]{\ensuremath{\big|#1\rangle}}
\newcommand{\bra}[1]{\ensuremath{\langle#1\big|}}
\newcommand{\be}{\begin{equation}}
\newcommand{\ee}{\end{equation}}
\newcommand{\DBCS}{\ensuremath{\Delta_\mathrm{BCS}}}
\newcommand{\Dsp}{\ensuremath{\Delta_\mathrm{sp}}}
\newcommand{\oD}{\ensuremath{\omega_\mathrm{Debye}}}
\newcommand{\dphi}{\varphi}

\newcommand{\Eq}[1]{Eq.~(\ref{#1})}
\newcommand{\eq}[1]{Eq.~(\ref{#1})}

\newcommand{\e}{\ensuremath{\mathrm{e}}}

\newcommand{\pdag}{{\phantom{\dagger}}}

\begin{document}

\title{Josephson effect between superconducting nanograins with discrete energy levels}

\author{Dominique Gobert}
\email{gobert@lmu.de},
\author{Ulrich Schollw\"ock}
\author{Jan von Delft}
\affiliation{Sektion Physik and Center for NanoScience,
  Ludwigs-Maximilians-Universit\"at M\"unchen, \\
 Theresienstr.\ 37, D-80333 M\"unchen, Germany}

\date{\today}

\begin{abstract}
{ We investigate} the Josephson effect between two coupled
  superconductors, { coupled by the tunneling of pairs of
  electrons},  in the regime that their energy level spacing is
  comparable to the bulk superconducting gap, but neglecting any
  charging effects. In this regime, BCS
  theory is not valid, and the notion of a superconducting order
  parameter with a well-defined phase is inapplicable.  Using the
  density matrix renormalization group, we calculate the ground state
  of the two coupled superconductors and extract the Josephson energy.
  { The Josephson energy is found to display a reentrant behavior
    (decrease followed by increase) as a function of increasing level
    spacing.} For weak Josephson coupling, a tight-binding
  approximation is introduced, which illustrates the { physical
    mechanism underlying this reentrance} in a transparent way.  The
  DMRG method is also applied to two strongly coupled superconductors
  and allows a detailed examination of the limits of validity of the
  tight-binding model.
\end{abstract}
\maketitle
\section{Introduction}
\label{Intro}

The Josephson effect, i.e.\ the flow of a zero-voltage current between two weakly
coupled superconductors, with a sign and amplitude that depends on the
difference of the phases of their respective order parameter, can be
regarded as one of the most striking illustrations of phase coherent
behaviour in a macroscopic system and as
one of the hallmarks of superconductivity.
Although the Josephson effect is in general well understood, there 
is still a regime in which it has not yet been studied in detail: 
superconductors that are so small that the
discrete nature of their energy levels becomes important.
In this regime,
the theory of Bardeen, Cooper and Schrieffer (BCS),
which the quantitative understanding of the Josephson effect has been
based on, is not applicable.

This is because one of the underlying assumptions in standard BCS
theory is the presence of a (quasi-) continuous energy band.  As
Anderson first pointed out\cite{Anderson59}, BCS theory is not
consistent anymore once the superconductor is so small that the mean
level spacing $d$ is of the order of the superconducting gap
$\Delta_\mathrm{BCS}$: { According to BCS-theory, the} dominant
contribution to pairing correlations comes from levels within a range
of order $\DBCS$ around the Fermi surface, but there are no levels
left within this range when $d > \DBCS$.

When it became possible to reach this regime experimentally
by doing transport measurements on superconducting grains
with a diameter of only a few nanometers\cite{BlackTinkham96,DelftRalph01}, 
interest was spurred in a description of the {  pair-correlated} state
that is also valid for $d > \Delta_\mathrm{BCS}$.
It turned out that the BCS interaction in
this regime had already been extensively studied in
the context of nuclear physics, where an exact solution of the reduced
BCS Hamiltonian with discrete energy levels
had been found in 1964 by Richardson\cite{RichardsonSherman64}.

Using this solution, it was possible to explore in detail the
breakdown of BCS theory as $d$ increases.  Several surprising insights
were gained, one of which being that BCS theory already becomes
unreliable when $d \geq \Delta_\mathrm{BCS}^2 /
\omega_\mathrm{Debye}$, in other words, long before the Anderson
criterion is met\cite{SchechterDelft01}.  The underlying reason is
that in this regime, BCS theory underestimates the contribution of the
so-called { ``far or distant levels'', i.e.\ energy} levels farther
away than $\Delta_\mathrm{BCS}$ from the Fermi surface.  If the
contribution of these levels is properly accounted for, remnants of
superconductivity turn out to persist even for $d >
\Delta_\mathrm{BCS}$. 
{ Indeed, the} recent experiments on small superconducting grains
\cite{BlackTinkham96,DelftRalph01} { indirectly} confirmed these
results, in the sense that { even for level spacings as large as $d
  \sim \DBCS$, they observed an even-odd effect as a clear indication
  of remaining superconducting correlations.}

Another issue that arises for small superconductors is that the
superconducting phase $\phi$ is not { well-defined:} When the mean
number of electron pairs $\langle N \rangle$ is so small that
fluctuations around $\langle N \rangle$ in the grand canonical
ensemble are not negligible anymore, $N$ has to be treated as fixed.
As a consequence, due to the uncertainty relation $[N, \phi] = i$,
the notion of an order parameter with a { well}-defined phase loses
its meaning.

Therefore, a very natural question arises: What is the fate of the Josephson
effect between two small superconducting grains, in a regime where
BCS theory breaks down, and where the notion of a
superconducting phase variable is no longer valid?

In this paper, we examine this question in detail by studying
  two pair-correlated grains, coupled by a tunneling term that allows
  pairs of electrons to tunnel between the grains. We study this
  system using the density-matrix renormalization group (DMRG), a
  powerful numerical approach applicable to strongly correlated
  systems, which has already proven to be useful for calculating
  the properties of single superconducting
  grains\cite{SierraDukelsky00,DukelskySierra99}.  We here use it to
  calculate the ground state of two coupled grains and to extract the
  Josephson energy.  For weak Josephson coupling, we also perform a
  tight-binding approximation and compare its results to those of the
  DMRG calculation for two coupled grains.

  We identify two competing effects due to the discreteness of the
  energy levels: Somewhat surprisingly, the Josephson energy is {
    found to be} enhanced for large level spacing due to the
  contribution of a single energy level.  At intermediate level
  spacing, a kinetic energy term dominates, which suppresses the
  Josephson energy.  The competition of these effects leads to a
  surprising reentrant behavior { (decrease followed by increase)
    of the Josephson energy as a function of increasing level
    spacing.} In the limit of vanishing level spacing, the BCS result
  is recovered.
  
  At this { point}, we should mention an important restriction on
  our analysis: In the regime of small superconductors that we are
  interested in, the charging energy for an electron pair to tunnel
  between the two superconductors can become huge, easily of the order
  of a few hundred Kelvin in the experiments of
  \cite{DelftRalph01}.  As will be explained in some detail in
  subsection \ref{tight_discussion}, the dominant effect of the
  charging energy is to suppress tunneling events { altogether} and
  thereby to destroy the Josephson effect. {  However, the interest of
  the present paper is to study the effects due to the discrete
  spacing of the energy levels rather than that of charging effects,
  which have been thoroughly examined already 
  \cite{LikharevAverin91,MatveevShekhter93,IansitiLobb89}.  
  Therefore, we set the charging
  energy to zero in this paper.
  
  To experimentally realize the no-charging-energy model studied here,
  one needs systems for which the mean level spacing is larger than
  the charging energy. In principle, it is possible to reduce the
  charging energy of isolated grains, e.g. by using a pancake-shaped
  grain geometry in order to increase the inter-grain capacitance
  area, or by embedding the grains in a strong dielectric medium.  ---
  A more radical and at this point purely speculative way of studying
  Josephson physics in the absence of charging effects would be} to use
  uncharged particles instead of electrons, e.g.\ a degenerate Fermi
  gas of charge-neutral cold atoms in a double-well trapping
  potential.  Although a ``superconducting'' phase for cold neutral
  fermionic atoms has not yet been observed, there are predictions
  that this should be possible\cite{HofstetterLukin02}.  Once this has
  been achieved, a natural next step would be to study the Josephson
  effect in this system, for which the charging energy would indeed be
  zero.

  The outline of the paper is as follows: In Section II, we review the
  theory of the Josephson effect in a way that is also applicable for
  small superconductors, for which standard BCS theory is not
  applicable, and we give a definition of the Josephson energy
  independently of a superconducting phase variable.  Section III
  contains a brief introduction to the DMRG method and its application
  to the system of two coupled superconductors.  Finally, in section
  IV we { present and discuss} the results of our calculation.

\section{\label{theory_section} Josephson effect for weakly coupled superconductors: Theory}

In this section, we review some standard results of the theory of
  the Josephson effect and explain their relation to the
  pair-tunneling models to be used below. Our discussion of the
  Josephson effect is restricted to weak coupling
  between superconductors, such that perturbation theory in the
  coupling can be applied. We are careful, however, to formulate the
  Josephson effect in such a way that a generalization beyond
  perturbation theory is possible; this is done in the last subsection
  \ref{strong_subsec}.

The physical assumption underlying perturbation theory is that 
the tunnel coupling between the superconductors is so weak that
it is energetically not favorable to create excited
states  with broken electron pairs in the individual grains.
Therefore, the low-energy states of the coupled system
will not contain any of these excitations,
which will only be present as virtual states in
perturbation theory.
In more quantitative terms, the weak-coupling condition is $E_J^0 \ll
\Delta_\mathrm{sp}$, where $\Delta_\mathrm{sp}$ is the lowest energy
of a pair-breaking excitation, and the Josephson tunneling matrix
element $E_J^0$ is defined in \eq{EJ_0} below.

We are also careful to formulate our discussion of the Josephson
effect independently of the notion of a superconducting phase
variable, such that it remains valid in the regime of small
superconductors.  The material in this section is mostly not new and
has been discussed in one way or the other previously
\cite{Ferrell88,deGennes99}, but we feel it is worth presenting it in a way that
makes the { ensuing} application to small grains evident.

\subsection{\label{phase_subsec} Josephson effect as a phase dependent delocalization energy}

In the grand canonical ensemble, the phase of a superconductor $\phi$
{ can be} defined via the action of the pair annihilation
operator\cite{ref:timereversed} 
$b_i = c_{i\uparrow} c_{i\downarrow}$, and the
state $\ket{\phi}$ is said to have a phase $\phi$ if \be
\label{phi_def}
\bra{\phi} b_i \ket{\phi} \sim e^{i \phi},
\ee
$\phi$ being independent of the state $i$ (this is
the case for the ground state of a superconductor). 
A familiar example is the well-known BCS ansatz wave function
$\ket{\phi} = \prod_i (u_i + v_i e^{i\phi} b^\dagger_i) \ket{0}$, 
where $u_i$ and $v_i$ are real.
\Eq{phi_def} implies that a state with definite
phase $\phi$ must be a superposition of { many} states $\ket{N}$, 
{ each of which has a } fixed number $N$ of electron pairs:
\begin{equation}
\label{phi_state}
\ket{\phi} = \sum_{N \geq 0} C_N \e^{i N \phi} \ket{N},
\end{equation}
subject to the condition that $\bra{N}b_i\ket{N+1}$ is real, and
with real coefficients $C_N$.


{ In the canonical ensemble, however, where the number of electron
  pairs $N$ is fixed,} the expectation value (\ref{phi_def}) vanishes,
and the notion of a superconducting phase $\phi$ is obviously not
valid.  Nevertheless, the concept of a phase \emph{difference} $\dphi$
between two coupled superconductors (``left'' and ``right'', say) is
still applicable, because the number of electron pairs on each
individual superconductor need not be definite as long as the total
number on both superconductors is fixed.  In analogy to \eq{phi_def},
$\dphi$ can, then, be defined as \be
\label{dphi_def}
\bra{\dphi} b^{\phantom{\dagger}}_{r} b^\dagger_{l} \ket{\dphi} \sim e^{i \dphi}.
\ee
Here, the operators $b_l$ and $b_r$  refer to energy levels $l$,
$r$ of the left and right superconductors, respectively.
As in \eq{phi_def}, one has to assume that the phase in \eq{dphi_def}
is independent of the levels $l$ and $r$ for $\dphi$ to be
well-defined.

An example of a state with definite phase difference $\dphi$ is, in
analogy to \eq{phi_state},
\begin{equation} 
\label{dphi_state}
\ket{\dphi}  
=  
\sum_{\nu = - N/2}^{N/2} C_\nu e^{i \nu \dphi}
\ket{\nu}.
\end{equation} 
with real coefficients $C_\nu$.
Here, the states $\ket{\nu}$ denotes arbitrary states with $N/2 - \nu$ 
pairs on the left and $N/2 + \nu$ pairs on the right superconductor,
subject to the condition that 
$\bra{\nu} b^{\phantom{\dagger}}_{r} b^\dagger_{l} \ket{\nu+1}$
is real.

We will be only interested in situations for which $| \nu  \rangle$
has the form
\be 
\label{nu_state}
\ket{\nu} = \ket{N/2 - \nu}_L \otimes \ket{N/2 + \nu}_R,
\ee
where $\ket{n}_{L,R}$ are the superconducting ground states 
of the isolated $L$- (``left'') or $R$- (``right'') superconductors,
{ each containing  a definite number of pairs, $n$}.
These states can always be chosen to satisfy the above reality condition.
 

As was pointed out by Josephson, the presence of a phase difference
$\dphi$ as in \eq{dphi_def} has observable consequences when two
superconductors are coupled: In particular, { for weak coupling the
  coherent tunneling of pairs induces a zero-voltage current,} 
\be
\label{I_J}
I = I_J \sin \dphi,
\ee
that explicitly depends on $\dphi$.
As is well known \cite{deGennes99,Tinkham96}, the Josephson current
can, via the relation  
\be
\label{I_from_E}
I =  (2e / \hbar) \partial E / \partial (\dphi),
\ee
also be interpreted as a dependence of the total energy 
$E$ on the phase difference $\dphi$.
For example, we expect \eq{I_J} to follow from the
energy-phase relation
\be
\label{E_phi}
E(\dphi) =  \mathrm{const} - E_J \cos \dphi, \;\; E_J = (\hbar / 2e) I_J
\ee
which we will derive explicitly in \ref{tight_discussion} in the limit
$d \rightarrow 0$.
A more general definition of $E_J$, consistent with \eq{E_phi},
will be given in subsection \ref{strong_subsec}, where we associate $E_J$ with
the energy gain in the ground state (i.e. $\dphi = 0$) due to
the coherent tunneling of electron pairs.

The Josephson energy $E_J$ sets the energy scale relevant for the Josephson effect:
It is a delocalization energy that characterizes the
coupling of two materials, their tendency to 
have the same phase and the maximum supercurrent 
$I_J = (2e / \hbar) E_J$ 
that can flow between them.

\subsection{Pair tunneling Hamiltonian}

Only processes that depend on the relative phase $\dphi$ are relevant
for the Josephson effect, as is illustrated in \eq{I_from_E}.  Because
of \eq{dphi_def}, such processes require the coherent tunneling of
electron pairs; therefore, they have to be treated at least in second
order in the tunneling of single electrons.  The main goal of this
subsection will be to derive an effective pair-tunneling Hamiltonian,
\eq{H_J} below, that arises at this order.

Consider two superconductors $L$ and $R$ (left and right), each having
equally spaced energy levels with level spacing $d$, 
and each with a reduced BCS interaction with (dimensionless) coupling
constant $\lambda$:
\begin{equation}
\label{H_LR}
H_L = \sum_{l \sigma} \epsilon_l c_{l \sigma}^\dagger c^\pdag_{l \sigma} 
    - \lambda d \sum_{l l'}  
    c_{l \downarrow}^\dagger c_{l \uparrow}^\dagger 
    c^\pdag_{l' \uparrow} c^\pdag_{l' \downarrow}, \; \;
H_R \; \mathrm{similarly},
\end{equation}
where $\epsilon_l = l\, d$ is the bare energy of level $l$, $\sigma =
\uparrow, \downarrow$ is the spin, and the sums are over all energy
levels closer to the Fermi surface than the Debye energy $\oD$.

Let $L$ and $R$ be coupled by single electron tunneling with constant
tunneling matrix element $t$,
\begin{equation}
\label{H_1e}
H_{1e} = -t d \sum_{l r \sigma} c_{l \sigma}^\dagger c_{r \sigma}^\pdag + h.c.
\end{equation}
The coupling (\ref{H_1e}) lowers the total energy by generating states such 
as (\ref{dphi_state}), that superimpose
different numbers of electrons on each superconductor.
{ For simplicity, we assume the sum in \eq{H_1e} to be cut off at
  $\oD$ in all numerical calculations below.}

To second order in $H_{1e}$, the tunneling processes can be described by the
effective tunneling Hamiltonian
\be
\label{H_2}
H_2 = - \sum_{r l \sigma \nu} \frac{H_{1e} \ket{r l \sigma \nu} \bra{r
    l \sigma \nu} H_{1e}} {E_{r l \nu}}, 
\ee 
acting on the space
spanned by the states $\ket{\nu}$, defined in \eq{nu_state}.  
{ The sum in \eq{H_2} runs over { all possible intermediate
    states} $\ket{r l \sigma \nu}$ that can be reached by removing a
  single $(r \sigma)$-electron from state $|N/2 - \nu \rangle_R$ and
  adding a single $(l \sigma)$-electron to state $|N/2 + \nu
  \rangle_L$.  $E_{r l \nu}$ is the corresponding excitation energy
  relative to the energy of the state $\ket{\nu}$.  We assume,
  however,} that for given $r,l,\sigma,\nu$, all states except the one
with the lowest energy give a { negligibly small contribution to
  the sum,} because of the following argument: In the BCS limit, which
is valid for large enough values of $\lambda$, all excited states are
described by the quasiparticle operators \cite{Tinkham96} 
\be
\gamma^\dagger_{(e) \sigma i} = u_i c^\dagger_{i \sigma} \mp v_i P^\dagger c^\pdag_{i (-\sigma)},
  \; \;
\gamma^\dagger_{(h) \sigma i} = u_i P c^\dagger_{i \sigma} \mp v_i c^\pdag_{i (-\sigma)},
\ee where
$P^\dagger$ is an operator that creates an additional pair.  In this
limit, it is easy to see that only the lowest energy state $\ket{r l
  \sigma \nu} = \gamma^\dagger_{(e) \sigma l} \gamma^\dagger_{(h)
  (-\sigma) r} \ket{\nu}$ gives a contribution to \eq{H_2}, whereas
all other intermediate states have a vanishing overlap with $H_{1e}
\ket{\nu}$.  This is also the case for $\lambda = 0$, where
$\gamma^\dagger_{(e) \sigma l} \gamma^\dagger_{(h) (-\sigma) r} =
c^\dagger_{\sigma l} c^\pdag_{\sigma r}$.  For intermediate values of
$\lambda$, no simple argument can be made; we expect, however, that
still the state with the lowest energy will give the dominant
contribution.

The energy $E_{r l \nu}$ is given by the
collective excitation energies $E_r + E_l$, arising from the fact that levels
$r$ and $l$ are singly occupied. In general, it will also include a
$\nu$ dependent contribution from charging energy due to the electron
tunneling, see \cite{MatveevShekhter93}, but we chose to consider
only situations in which these can be neglected.

In \eq{H_2}, two kinds of tunneling terms are present:
on the one hand, terms proportional to $b^\dagger_l b^\pdag_r$
or to  $b^\dagger_r b^\pdag_l$ that describe
coherent pair tunneling,
on the other hand, single electron terms proportional to
$c^\pdag_{r \sigma} c^\dagger_{r \sigma} c^\dagger_{l -\sigma} c^\pdag_{l
  -\sigma}$
that describe the tunneling of a single electron from $l$ to $r$ and
back.
When the former terms are applied to a state $\ket{\dphi}$, defined in 
\eq{dphi_state}, they produce a phase dependent energy shift.
In contrast, the latter terms only lead to a phase-independent
energy shift, which is  irrelevant for the Josephson effect.
For this reason, the single electron terms can be omitted from
the Hamiltonian (\ref{H_2}), as long as only phase dependent processes
are of interest\cite{Ferrell88}. 
Then, one finally arrives at the pair tunneling Hamiltonian
\be
\label{H_J}
H_J = 
- 2 \sum_{rl} \frac{\gamma d^2}{E_r + E_l}
  ( b^\dagger_{r} b_{l} + h.c.),
\ee
with $\gamma = t^2$.
We shall use for the excitation energies their BCS values, $E_{r,l} =
\sqrt{\DBCS^2 + \epsilon_{r,l}^2}$.

\subsection{\label{tightbindingmodel} Tight-binding model}

In the space spanned by all states without any pair-breaking excitations, 
i.e.~all states of the form $\ket{\nu}$ defined in \eq{nu_state}, 
the Hamiltonian $H = H_L + H_R + H_J$ looks like a tight-binding
Hamiltonian:
\begin{equation}
\label{Htightbinding}
H = \left(
\begin{array}{lllll}
E(\nu)   & -E^0_J/2    &  0     & \hdots &  \\
-E^0_J/2 & E(\nu)      & -E^0_J/2 & 0 &\hdots  \\
 0  & -E^0_J/2 & & & \\
\vdots & 0 & & & \\
 &\vdots & & &
\end{array} 
\right),
\end{equation}
where 
\begin{equation}
\label{EJ_0}
E^0_J/2 = -\bra{\nu} H_J \ket{\nu+1}, 
\ee
\be
E(\nu) =  \bra{\nu} (H_L + H_R) \ket{\nu}.
\end{equation}

Much fewer electrons will tunnel than are present on each
superconductor, $\nu \ll N$. Therefore,
the off-diagonal elements $E_J^0$ can be taken to be independent of $\nu$.
The diagonal elements are given by
\be
\label{E_kin}
E(\nu) =  \mathrm{const} + 2d (\nu - \nu_0)^2,
\ee
which has the form of an effective charging energy term.
This is because changing $\nu$ by one is equivalent to shifting the
relative chemical potential between the grains
by the amount $2d$, except for the outmost 
energy level (i.e.\ the level closest to the cutoff at $\oD$), which can 
be neglected if $\nu \ll N$.



\subsection{\label{tight_discussion}Discussion of the tight-binding model}

In this subsection, we first discuss the above tight
binding model (\ref{Htightbinding}) in the limit $d \rightarrow 0$
and check that it is
consistent with the well-known result of Ambegaokar and Baratoff \cite{AmbegaokarBaratoff63}.
Then, we draw attention to what changes will occur as the superconductors
become smaller.

For $d \rightarrow 0$,  the diagonal
elements (\ref{E_kin}) of $H_J$ [Eq.~(\ref{Htightbinding})] are 
{ independent of $\nu$. }
Also, BCS theory is
valid, so the off-diagonal elements (\ref{EJ_0}) are given by 
\begin{equation}
\label{EJ_large}
E_J^0
= 
\sum_{lr} \frac{\DBCS^2 t^2 d^2}{E_l E_r (E_l + E_r)}
= 
\DBCS t^2 \pi^2.
\end{equation}
In the left equality of \eq{EJ_large}, the BCS expression for
the matrix elements
$\bra{ \nu }  b^\pdag_{l} b^\dagger_{r} \ket{\nu+1} = v_l u_l v_r u_r =
 \DBCS^2 / (2 E_l E_r)$ has been used.
For the right equality, the sum has been replaced by an integral, 
$\sum_{ij} = \int_{-\infty}^{\infty}\frac{d\epsilon_1 \; d\epsilon_2}{d^2}$.
No harm is done extending the integral range beyond $\omega_\mathrm{Debye}$
to infinity, because it is naturally cut off at the scale
$\DBCS$ anyway, assumed to be much smaller than $\omega_\mathrm{Debye}$.

The energy eigenstates of (\ref{Htightbinding}) then have the form of
\eq{dphi_state} with constant coefficients $C_\nu$.  As anticipated in
\eq{E_phi}, they correspond to an energy $E(\dphi) = \mathrm{const} -
E_J^0 \cos \dphi$, and therefore we can identify 
\be
\label{EJ_BCS}
E_J = E_J^0 = \Delta t^2 \pi^2 = \frac{\pi \hbar \Delta_{\rm BCS}}{4 e^2 R_N} \; .  \ee { The last equality expresses $E_J$ in terms
  of the normal-state conductance $R_N^{-1} = (4\pi e^2 / \hbar) t^2$,
  and agrees} with the well-known Ambegaokar-Baratoff
formula\cite{AmbegaokarBaratoff63} at zero temperature.

Now we turn to the question what happens when the superconductors
enter the regime $d \geq \DBCS^2 / \oD$, in which the BCS Ansatz
wavefunction becomes inappropriate \cite{SchechterDelft01}.  The
transition to this regime is straightforward now, because the
tight-binding model itself remains valid: The diagonal and
off-diagonal matrix elements $E(\nu)$ and $E_J^0$ of the tight-binding
Hamiltonian (\ref{Htightbinding}), defined in \eq{E_kin} and
(\ref{EJ_0}), will no longer be given by the BCS expression, but will
have to be evaluated using the exact ground state wave function:  The
effect of the discrete level spacing on the diagonal elements
$E(\nu)$, given by \eq{E_kin}, will be to lift the degeneracy among
them, thereby suppressing pair tunneling.

The off-diagonal elements $E_J^0$ will change with respect to their
BCS value (\ref{EJ_large}) due to several effects when the superconductors become
small:
(i) The excitation energies $E_{r l \nu}$ in \eq{H_J} will
include a term from the charging energy in the intermediate state, as
studied in \cite{MatveevShekhter93}.
We neglect this effect, because we have chosen not to study charging effects at all.
Furthermore, the change in superconducting correlations due to the finite
level spacing will affect both
(ii) the excitation energies $E_r$ and $E_l$ in \eq{H_J} 
(which we, however, replace with their BCS value) and 
(iii) the matrix
elements $\bra{\nu} b_l b^\dagger_r \ket{\nu+1}$ that enter $E_J^0$.
Finally, (iv) the shift of the Fermi level between the states $\ket{\nu}$ and
$\ket{\nu+1}$ will also change the matrix elements $E_J^0$, as is
explained in section \ref{results_tightbind} below.


As it turns out (see  section \ref{results_tightbind} below), 
$E_J^0$ \emph{increases} with increasing level spacing $d$, mainly  due
to the effect (iv).
Once  $E_J^0$ becomes comparable to $\Delta_\mathrm{sp}$, the
lowest pair breaking (``single-particle'') excitation energy, the
superconductors can no longer be considered as weakly 
coupled, and the tight-binding model itself loses its validity.

\subsection{The effect of charging energy}

As mentioned in the introduction, the Coulomb charging
energy plays an important role in small superconductors. Although
we shall neglect it in the remainder of this paper,
here we present a brief qualitative discussion of its { main} effects. 
The charging energy $E_C = (2e)^2/C$, $C$ being the inter-grain
capacitance, is the energy cost for tunneling an electron pair from one grain
to the other. { It} introduces an additional term in (\ref{Htightbinding}),
$E(\nu) = E_C (\nu - \nu_0)^2$.
$E_C$ can become huge in the small grain limit
and essentially destroys the Josephson effect, since it
suppresses pair tunneling. 

Even if a gate is used to make two states $\ket{\nu}$ and $\ket{\nu+1}$ 
degenerate by a suitable  choice of the gate voltage (i.e.\ $\nu_0
= 1/2$ plus an integer),
such that at least one pair can still tunnel between the grains at no energy cost, the
charging energy might nevertheless destroy the Josephson effect altogether:
It may cause one electron pair to break into two unpaired electrons,
one on each grain, if the associated lowering of the charging energy exceeds the
energy necessary to form a pair-breaking excitation.

An order-of-magnitude estimate shows that this actually happens in the
regime that the level spacing is important, if no measures are taken
to reduce the charging energy: (i) As explained above, the charging
energy must be smaller than the lowest energy of a pair breaking
excitation, $E_C < \Delta_\mathrm{sp}$, such that no pair breaking
excitations occur.  (ii) $\Delta_\mathrm{BCS} <
\sqrt{\omega_\mathrm{Debye} \, d}$ must be satisfied if the grains are
to be small enough so that deviations from BCS become important (the
'weak' criterion in \cite{SchechterDelft01}).  (iii) { For
  the present purpose of constructing an order-of-magnitude estimate,
  we take $\Delta_\mathrm{sp} \sim \Delta_\mathrm{BCS}$, although
  these two energy scales may not be identical in the small-grain
  limit \cite{SchechterDelft01}. (They differ, for example, by
  a factor of up to two for the parameter range shown in Fig.\ 1 of
  \cite{SchechterDelft01}.)}

Putting (i) to (iii) together, the inequality
\begin{equation}
\label{charge_breakdown}
E_C < \sqrt{\omega_\mathrm{Debye} \, d},
\end{equation}
independent of $\lambda$, has to be satisfied.

{ Let us now explore what this implies for real Aluminium grains:}
If the inter-grain capacitance is modelled by an Aluminium oxide
($\epsilon \approx 8$) layer of thickness $15$ \AA and area $\pi r^2$,
then $E_C \approx 0.8 \; \mathrm{eV} (r/\mathrm{nm})^{-2}$.  (A
smaller thickness $D$ in principle linearly decreases the charging
energy, but at the same time, the inter-grain coupling $t$ is
exponentially increased\cite{KnorrLeslie73}, $t^2 \propto \exp[ -
D/(0.54\mathrm{\AA})]$.  Since at a thickness of less than $\sim
15$ \AA, the grains are so strongly coupled that thay can no
longer be considered as distinct, this distance seems to be a
realistic order-of-magnitude lower bound for $D$.)  Using the Debye
energy $\omega_\mathrm{Debye} = 35\;\mathrm{meV}$ for Aluminum, we
obtain $\sqrt{\oD \cdot d} = 0.054 \; \mathrm{eV}
(r/\mathrm{nm})^{-3/2}$, and \eq{charge_breakdown} implies $r > 250 \;
\mathrm{nm}$.  At such a large size, Aluminum is well in the BCS
regime.  According to criterion (ii) above, { deviations from the
  BCS approach for a grain of that size would be observable only for a
  material with $\Delta_\mathrm{BCS} < 10^{-5} \; \mathrm{eV}$, an
  order of magnitude less than Al.}

As an example, we consider the experiments of Nakamura et al
\cite{NakamuraTsai99}, which use a superconducting island with $\Delta
\approx 230 \; \mu\mathrm{eV}$ and $E_C \approx 117 \;
\mu\mathrm{eV}$.  These islands are evidently so small that they are
quite close (up to a factor of 2) to the regime where the charging
energy would begin to suppress pair tunneling by favoring
single-particle excitations.  Nevertheless, their islands are still
large enough to be well described by BCS theory.

However, as mentioned in the introduction, the interest of this paper 
is to study the effects due to the discrete 
spacing of the energy levels, since the charging
effects have already been discussed previously
\cite{LikharevAverin91,MatveevShekhter93,IansitiLobb89}.
Therefore, we henceforth set the charging energy to zero.

\subsection{\label{strong_subsec}Generalization to strong coupling}

In the weak coupling limit, we have defined the Josephson energy via the
part of the energy (\ref{E_phi}) that depends on $\dphi$.
However, \eq{E_phi} is only valid for weak coupling
(i.e.\ in second order in the single electron tunneling).
We may equivalently define the Josephson energy as the maximally
possible energy lowering
due to coherent {\emph{pair}} tunneling, i.e.\ when single electron terms are
neglected as in the derivation of (\ref{H_J}):
\begin{equation}
  \label{EJ_def}
  E_J \equiv E_\mathrm{coupled} - E_\mathrm{uncoupled}.
\end{equation}
This definition agrees with the usual one
(\ref{E_phi}) in the weak-coupling regime, because the maximally
possible energy lowering occurs at phase difference $\dphi = 0$.
\eq{EJ_def} allows an extrapolation to strong coupling as well, and
therefore we will use it henceforth.

Unfortunately, the pair tunneling Hamiltonian (\ref{H_J}), being only
derived in  second order perturbation theory, loses its validity for
strong coupling; in general,
one would have to use the single-electron tunneling Hamiltonian
(\ref{H_1e}) in that case.
For simplicity, however, we choose
for our strong coupling analysis a somewhat different coupling term that
only includes pair tunneling,
\begin{equation}
\label{H_J_flat}
H_J' = -
\frac{\gamma d^2}{\DBCS} \sum_{rl}
  ( b^\dagger_{1 r} b_{2 l} + h.c.),
\end{equation}
and that differs from the pair tunneling Hamiltonian (\ref{H_J}) in
that the intermediate energy $E_r + E_l$ has been replaced by the
constant $\DBCS$.  Therefore, the Hamiltonian (\ref{H_J_flat}) and
(\ref{H_J}) are not equivalent.  It is nevertheless interesting to
study the Hamiltonian (\ref{H_J_flat}) for several reasons: Firstly,
it captures the essential physics of the Josephson effect in a simple
way: two superconductors coupled by a tunneling barrier that allows
for pair tunneling.  After all, it is the pair tunneling and not the
single electron tunneling that is at the heart of the Josephson
effect.
Secondly, for $\gamma d / \DBCS = \lambda$, the total Hamiltonian
looks just like one single superconductor, thus (\ref{H_J_flat}) is
able to describe the transition to { the strong-coupling regime
  where two superconductors} effectively become one.  { Thirdly, it
  is amenable to a rather straightforward treatment by the DMRG
  approach (in contrast, $H'_J$ of Eq.~(\ref{H_J}) would require much
  more numerical effort), which has the very significant advantage of
  yielding direct access to the regime of strong coupling between the
  two superconductors.} 

 At weak coupling, a tight-binding analysis
for (\ref{H_J_flat}) similar to the one that led to \eq{EJ_large} can
be performed.  In the large grain limit, one finds the Josephson
energy to be \be
\label{EJ_flat_BCS}
E_J^0 =  \frac{2 \gamma \DBCS}{\lambda^{2}},
\ee
independent of $d$. 
In other words, the Hamiltonian (\ref{H_J_flat}) has a well-defined 
continuum limit when $\gamma$ is held contant as $d \rightarrow 0$, as it should.

\section{\label{DMRG_sec}DMRG approach}

In the context of nuclear physics, Richardson found an exact solution
\cite{RichardsonSherman64} of the Hamiltonian (\ref{H_LR}) for a
single superconductor, that allows in principle to calculate all of its
eigenenergies and eigenstates.
Because the tight-binding calculation for weakly coupled
superconductors, as outlined in \ref{tightbindingmodel},
only needs the matrix elements (\ref{EJ_0}) between states of a single superconductor,
Richardson's solution is, in principle, sufficient for that case.

However, while the eigenenergies of (\ref{H_LR}) can be calculated
with only little numerical effort using Richardson's solution, the
computation time needed for the eigenstates and for matrix elements
like the ones in (\ref{EJ_0}) scales like $n!$ with the number of
energy levels $n$ in the system, making it effectively impossible to go
beyond, say, $n=12$ levels or so { (more precisely, only the number $n$
of energy levels between $E_\mathrm{Fermi}-\oD$ and
$E_\mathrm{Fermi}+\oD$ matters).}
For this reason, despite
there being an exact solution available, it is indispensable also for
the tight-binding model to have an alternative approach at hand that
is approximate, but manageable.  Moreover, for the strong coupling
analysis in { Sec.~\ref{strong_subsec}, that invokes the pair
tunneling term (\ref{H_J_flat}), Richardson's solution is not
applicable at all, so that the use of
a different approach becomes unavoidable.}

For these reasons, we have adopted an approach based on the density
matrix renormalization group (DMRG), whose power and efficiency for
{ dealing with pair-correlated nanograins has been}
demonstrated recently\cite{SierraDukelsky00,DukelskySierra99}.  We will use two kinds of DMRG calculations: A
single-grain DMRG for calculating the matrix elements (Sec.~\ref{EJ_0}) to
be used in the tight-binding model at weak coupling (cf.\
Sec.~\ref{tightbindingmodel}), and a two-grain DMRG for the case of strong
coupling (cf.\ Sec.~\ref{strong_subsec}).

In this section, we first discuss some general aspects of the DMRG
algorithm in energy space in \ref{DMRG_subsec}, leaving some of the
more technical issues for appendix \ref{DMRG_app}.  In
Sec.~\ref{perturb_DMRG}, we discuss the one-grain DMRG, and turn to
the dicussion of the two-grain DMRG in
Sec.~\ref{twograin_DMRG_subsec}.

\subsection{\label{DMRG_subsec}The DMRG method in energy space}

The DMRG in its usual implementation is a real-space renormalization
group method, and has been very successful for describing one dimensional
many-particle quantum systems, such as spin chains \cite{Peschel99}.
Usually, the Hilbert space for such systems is too
large to be diagonalized exactly on a computer.
The DMRG algorithm allows to keep only a reduced part of the Hilbert
space that is small enough to be tractable even on a desktop computer,
but still sufficient to describe one or several desired states, the
so-called target states (in our case, the ground state will be the
target state).
This is achieved by progressively increasing the chain size,
adding sites one at a time, while only a limited number
of states is kept at each step, those states being selected as the
most relevant ones for describing the target state(s) in a density
matrix analysis.

Although the DMRG is mostly limited to one dimensional systems, it can
be applied to three dimensional ones by using the energy axis as the
one dimensional ``system'', such that the bare energy levels play the
role of sites on a one dimensional chain.  This is not always useful,
because the interactions between these ``sites'' can be much messier
than between sites in real space, the latter being generally local.
Luckily, as will be seen, the BCS interaction is, although nonlocal,
simple enough for the DMRG algorithm to be applicable.

The DMRG builds up the system, starting from the low-lying energy states
around the Fermi surface, which are  the physically most important
ones, and successively adds levels lying further and further away from the
Fermi energy.
It should be noted that this is quite contrary to the way usual RG
calculations are performed, where high energy levels are integrated out, 
approaching the low energy states from above.
This allows these two complementary approaches to be simultaneously
applied:
As long as not all energy levels have yet been added to the system,
only the ones near the Fermi surface are explicitly included in the DMRG
calculation.
The other ones, which will be included only at later
steps, are meanwhile taken into account using a
renormalization of coupling constants (as introduced in eq.\ 
(43) of \cite{SierraDukelsky00}).

For this purpose, the following scheme turns out to be numerically
very efficient for renormalizing the coupling constants $\lambda$ and
$\gamma$: When { the $i$ levels closest to the Fermi energy} are
included, choose the coupling constants, say $\lambda_i$ and
$\gamma_i$, such that the BCS band gap $\Delta_i$ of the current
system equals the final value $\Delta_n$, where $n$ is the desired
final number of levels.  In the DMRG for a single grain,
$\Delta^{(1)}_i = i d / (2 \sinh(1/\lambda_i))$.  In the two-grain
DMRG, the band gap is given by $\Delta^{(2)}_i = i d / \sinh(1/
(\lambda_i + \gamma_i d / \Delta_n))$.  The latter is the the solution
of the BCS gap equation with two different interaction matrix elements
$-\lambda_i d$ and $-\gamma_i d^2 / \Delta^{(1)}_n$, as in
(\ref{H_LR}) and (\ref{H_J_flat}).  For large couplings, this scheme
turns out to be more efficient than a perturbative renormalization of
the coupling constants.  At weak couplings, for which perturbation
theory is expected to work, both approaches perform equally well.

Another drastic reduction of degrees of freedom occurs because in the
model we study, the energy levels that are occupied by a single
electron completely decouple from all the interaction terms (\ref{H_LR}),
(\ref{H_J}) and (\ref{H_J_flat}).
Because the creation of a singly occupied level is associated with the
energy $\Delta_\mathrm{sp}$ and therefore  energetically
unfavorable, there will be no singly occupied levels in the low-energy
sector of a superconductor, if one assumes the total number of
electrons to be even.
Due to these considerations, we can omit these levels from the
beginning, and consider only the case of empty or doubly occupied
energy levels \cite{DelftRalph01}.

Although the full Hilbert space is drastically reduced by the DMRG algorithm,
it produces excellent results.
In the case of the two-grain DMRG,
the accuracy can be checked by comparing the condensation energy from DMRG 
to the Richardson solution, which is available for two specific
values of the inter-grain coupling $\gamma$ in (\ref{H_J_flat}), namely for
\cite{LinksHibberd02} 
$\gamma d / \Delta = \lambda$ (which effectively describes one
single, larger superconductor)
and $\gamma = 0$ (two independent superconductors).
The results for the two-grain DMRG are shown in Fig.\ 
\ref{m_deltaE} and show the following features:
\begin{figure}
\epsfig{file=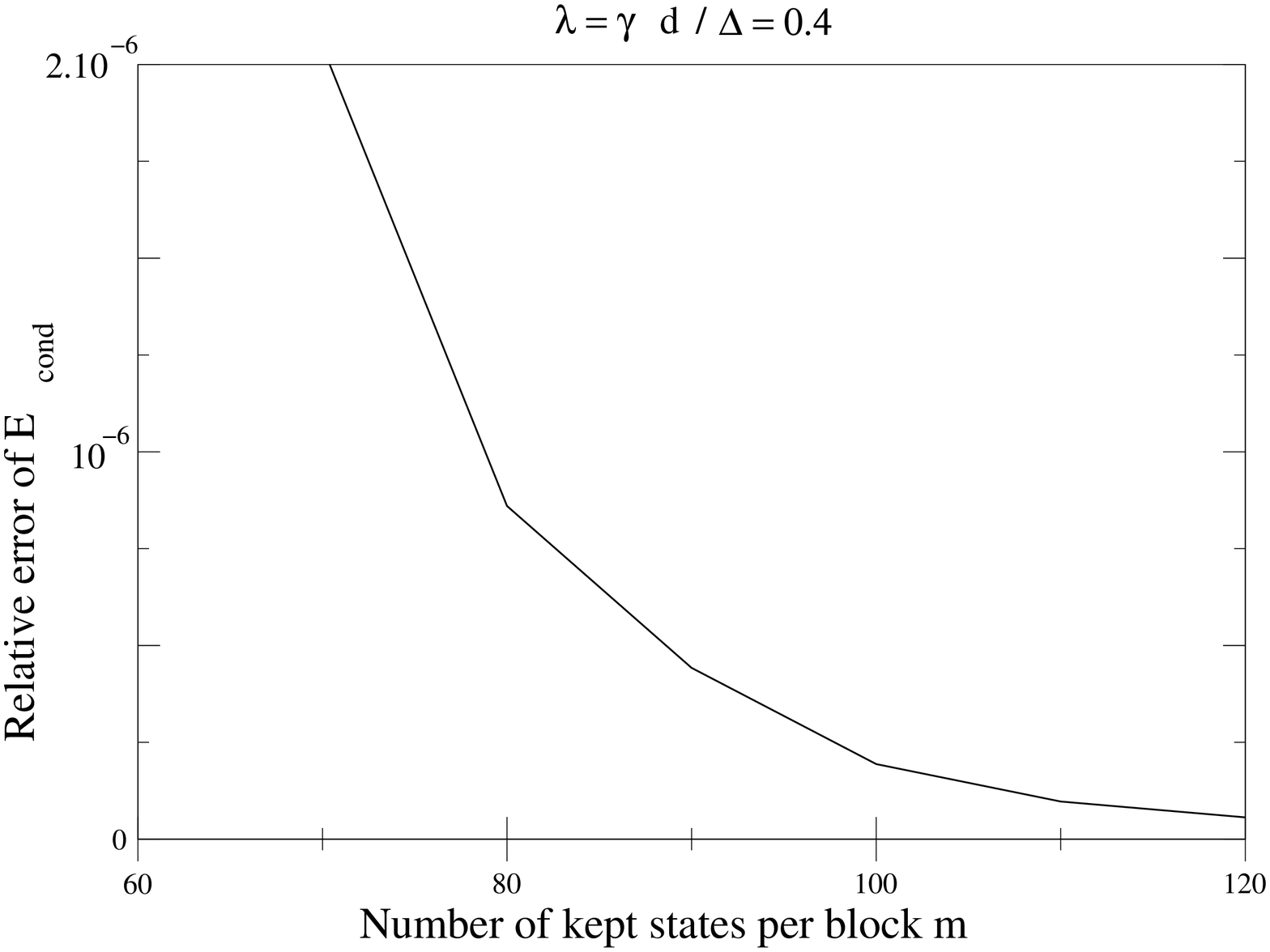, width=0.9\linewidth}
\epsfig{file=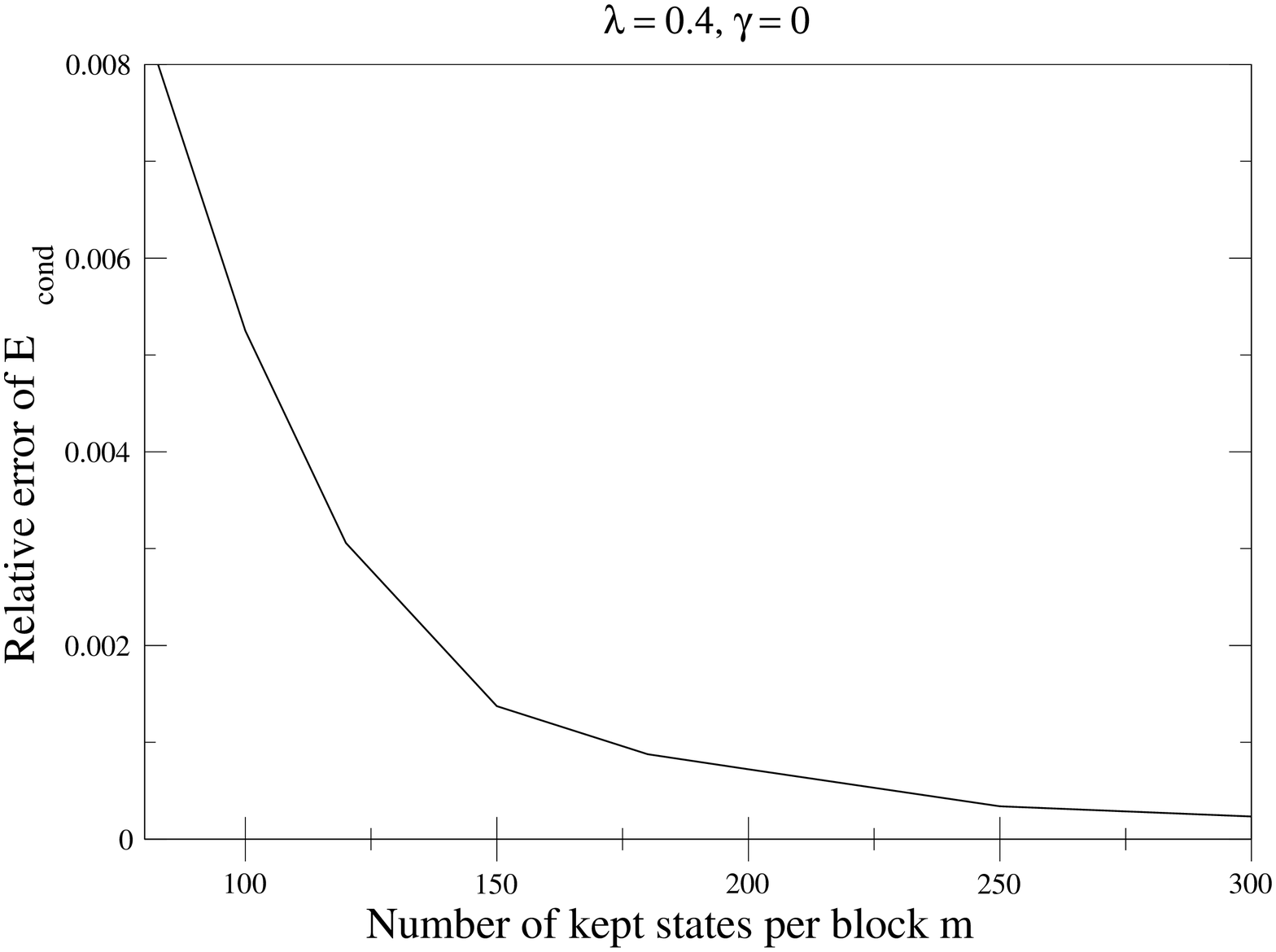, width=0.9\linewidth}
\caption{\label{m_deltaE} Relative error between the exact result
  from Richardson's solution
  and the two-grain DMRG at BCS coupling $\lambda = 0.4$, with $n = 100$ 
  energy levels per grain.
  The inter-grain coupling in the upper plot is
  $\gamma d / \Delta = \lambda$.
  In the lower plot, $\gamma =0$.
}
\end{figure}
(i)
 High precision at strong inter-grain coupling, with a relative
  error in the condensation energies of only $\sim 10^{-7}$ when
  $m=100$ states are kept.
(ii)
 Lower, but still sufficient precision for decoupled grains
  ($\gamma = 0$): $\sim 10^{-3}$ for $m=300$, for $n = 100$ energy levels.
  However, the algorithm fails at weak coupling when the number of
  energy levels $n$ becomes large ($N > 80-150$), see
  \ref{twograin_DMRG_subsec}. 
  In this case, a perturbative calculation (see \ref{perturb_DMRG})
  becomes necessary.

With the one-grain DMRG, the accuracy of case (i) is obtained.
As always in DMRG, the precision can be systematically improved by
increasing $m$.

\subsection{\label{perturb_DMRG}One-grain DMRG for tight-binding model}

If the grains are weakly coupled, the tight-binding approach can be
applied, based on the Hamiltonian (\ref{H_J}).
Here, the microscopic model only enters via the tunneling matrix elements  
$E_J^0$ (\ref{EJ_0}) of the tight-binding Hamiltonian
(\ref{Htightbinding}).  Although these can in principle be calculated
{ exactly using Richardson's solution, in practice} the DMRG
algorithm is much better suited for that task, as explained above.

Assuming $\nu \ll N$ and using \eq{nu_state}, the only matrix elements
needed for $E_J^0$ are $\bra{N/2+1} b_i^\dagger \ket{N/2}$ for all
values of $i$.  We evaluate these matrix elements using the DMRG
algorithm for one single grain, as introduced in {
  Refs.}~\cite{SierraDukelsky00,DukelskySierra99}.  This
requires the simultaneous knowledge of two ground states with
differrent pair occupation numbers, $\ket{N/2}$ and $\ket{N/2 + 1}$.
These states are constructed in a single run, as explained in 
appendix \ref{DMRG_app}.  Once these matrix elements have been
calculated, it is straightforward to diagonalize the tight-binding
Hamiltonian (\ref{Htightbinding}).

\subsection{\label{twograin_DMRG_subsec}Two-grain DMRG}

If the DMRG is directly applied to a system of two grains, the
regime of strong coupling can be explored, too.
For this purpose, we use the inter-grain coupling term (\ref{H_J_flat}), introduced in 
subsection \ref{strong_subsec}. 
The exact Richardson solution cannot be applied for this system
(except for the particular value of $\gamma = \lambda
d / \DBCS$, which has been used in subsection \ref{DMRG_subsec} for
checking the accuracy of the results).

Although the two-grain DMRG can cover the previously unaccessible
parameter region of strong coupling, it turns out to fail for too weak
inter-grain coupling, if the system is large (more than, say, 80-150
or so energy levels, depending on the other parameters).  The reason
is that the DMRG relies on correlations between the grains for being
able to effectively reduce the Hilbert space, and these correlations
vanish in the limit $\gamma \rightarrow 0$.  This can easily be seen
in the limiting case $\gamma = 0$, in which the two grains $L$ and $R$
are completely uncorrelated and can, each, be described by $m_1$
independent basis vectors $\ket{1}_L, ..., \ket{m_1}_L$ and
$\ket{1}_R, ..., \ket{m_1}_R$.  This implies that the $m$ basis
vectors retained are essentially product states  of the form
$\ket{i}_L \otimes \ket{j}_R$, and only an accuracy corresponding to
$m_1 = \sqrt{m}$ kept basis vectors per grain is attained.  If the
inter-grain coupling $\gamma$ is increased, correlations between
grains $L$ and $R$ quickly develop that allow to keep only a few
dominant ones of the product states, but for $\gamma = 0$, and also
for very small values of $\gamma$, each of these states is equally
important, making the DMRG highly inefficient.  That the DMRG still
works even for $\gamma = 0$ if only a few ($< 80-150$) energy levels
are considered, is due to the fact that in this case, the necessary
number of states to be kept per grain seems to be so low ($m_1 \approx
15$) that the ground state can still be reasonably well approximated.

To summarize, the two-grain DMRG works well for strongly correlated
systems, but produces unsatisfying results for the case of weak
inter-grain coupling.  However, this is the regime in which
perturbation theory can be used, as described before: thus, the
two-grain DMRG and perturbation theory are two complementary
approaches; their regimes of usefulness are illustrated in Fig.\ 
  \ref{tb_OK_fig} below.

\section{\label{results_sec}Results}

\subsection{ \label{results_tightbind}
From small to large grains: The effect of discrete energy levels
within the tight binding model}

\begin{figure}
  \epsfig{file=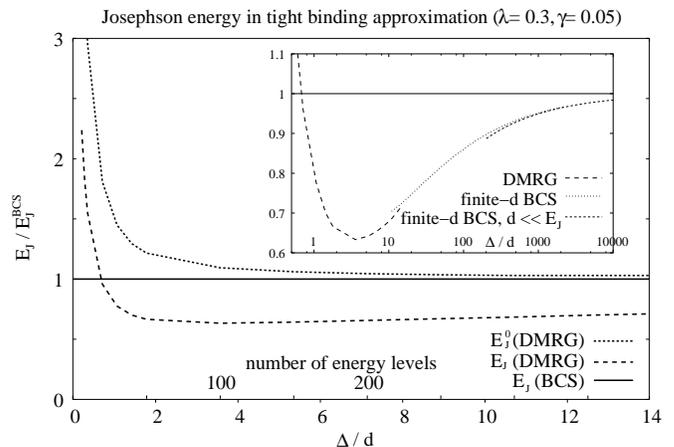, width=1\linewidth}
  \caption{\label{EJ_tight_fig} 
    The  Josephson energy $E_J$ in the tight-binding approximation, based on \eq{Htightbinding}, as
    a function of the grain size. $E_J$ is defined via \eq{EJ_def} as the
    additional energy gain due to coherent pair tunneling
    and is normalized to the BCS result $E_J^\mathrm{BCS}$ in \eq{EJ_BCS}.
    Compared to the off-diagonal matrix element $E_J^0$
    (dotted line), $E_J$ (dashed line)
    is reduced  by a factor of up to 2 due to the finite-size
    kinetic energy term $E(\nu)$ of \eq{E_kin}.
    The logarithmic plot in the inset shows how the
    BCS result of \eq{EJ_BCS} is recovered as $d \rightarrow 0$.}
\end{figure}

{ Fig.\ \ref{EJ_tight_fig} presents our results for} the Josephson
energy $E_J$, defined as the delocalization energy (\ref{EJ_def}) due
to pair tunneling, in the tight-binding approximation, { calculated
} using the coupling Hamiltonian given by \eq{Htightbinding}.  $E_J$
is plotted in units of the BCS result $E_J(BCS)$, given by
\eq{EJ_BCS}.  The dashed line in Fig.\ \ref{EJ_tight_fig} displays the
Josephson energy as a function of decreasing level spacing $d$,
i.e.\ of increasing grain size, characterized by the number
of discrete energy levels $n$ between $\varepsilon_F \pm \oD$.

While the level spacing $d$ is varied, the parameters
$\lambda$ and $\gamma$ in (\ref{H_J}) are held fixed (at the values
$\lambda = 0.3$, $\gamma = 0.05$), such that the BCS value in
\eq{EJ_large} of $E_J$ is independent of the grain size, and a
well-defined limit $d \rightarrow 0$ exists.

Within the tight-binding model, we observe two competing effects, to
be discussed in detail below, that influence the Josephson energy as
the level spacing $d$ increases (i.e.\ { moving toward} the left
side of Fig.\ \ref{EJ_tight_fig}): (i) On the one hand, the {
  finite-size} kinetic energy term $E(\nu)$ of \eq{E_kin} increases,
which tends to reduce the Josephson energy $E_J$; (ii) on the other
hand, the off-diagonal matrix element $E_J^0$ in \eq{Htightbinding}
increases (as shown in Fig.\ \ref{EJ_tight_fig}, dotted line), which
tends to increase $E_J$.  The combination of these two tendencies
leads to the reentrant behaviour seen in Fig.\ \ref{EJ_tight_fig},
{ particularly in the inset, 
with a remarkable}  {\emph{increase}} in $E_J$ when $d$ becomes
sufficiently large.

The kinetic term (i) was discussed in section \ref{tightbindingmodel}.
We chose $\nu_0$ in \eq{E_kin} as $\nu_0 = 1/2$, such that the two
lowest-lying states are degenerate and at least one pair can tunnel at
no energy cost between these states no matter how large $d$ is.  For
this case, { the total reduction of the Josephson energy due to the
  finite-size kinetic term amounts to a factor of at most 2,} even for
very large level spacing $d$.  This is because for $d \rightarrow
\infty$, { all but the lowest two states ``freeze out'', so that }
the tight-binding Hamiltonian (\ref{Htightbinding}) effectively
reduces to a two-level system for the states $\ket{\nu = 0}$ and
$\ket{\nu = 1}$, { whose tunnel splitting is $E_J^0/2$ (hence the
  reduction by a factor of 2).}  Nevertheless, the { reduction} in
Fig.\ \ref{EJ_tight_fig} is seen to be considerable even for fairly
large grains (still of order 20 \% for $\DBCS / d \sim 100$,
corresponding to $n \sim 3000$ levels), because it depends on the
ratio $d / E_J^0$, { where $E_J^0$ typically is} a small number
itself.  
For $ d \ll E_J$, the asymptotic behaviour $ E_J = E_J^0 ( 1
- \sqrt{2 d / E_J^0})$ (thin dashed line in the inset of Fig.\
\ref{EJ_tight_fig})
is found in analogy to the treatment of small charging  
energies in section 7.3 of \cite{Tinkham96},
by using an Ansatz wave function given by \eq{dphi_state} with $\dphi =
0$ and $C_\nu$ of Gaussian form. 
This Ansatz wave function turns out to be asymptotically correct for $d <<
E_J^0$ \cite{Tinkham96}.

\begin{figure}
  \epsfig{file=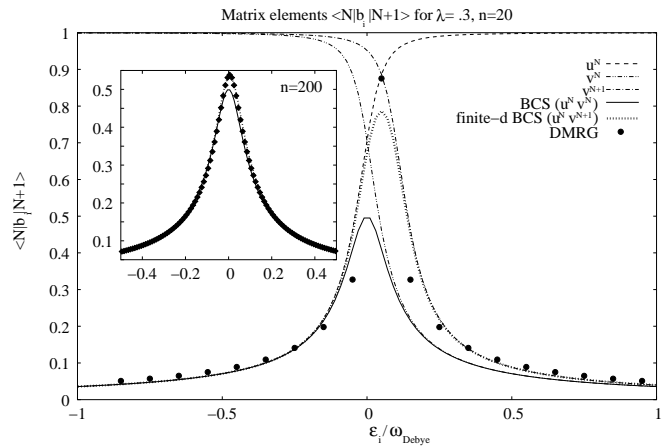,width=\linewidth}
  \caption{\label{uv_fig}
    Various approximations for the matrix element
    $\bra{N}b_i\ket{N+1}$ are compared.
    The product of the BCS coherence factors $u^N v^N$ (BCS, thin dashed
    line) is compared to $u^N v^{N+1}$ (finite-$d$ BCS, solid
    line) in a grain of $n=20$ levels.
    Because $v^{N+1}$ in the latter product is shifted to
    the right by an amount of $d$ with respect to $v^N$, the
    finite-$d$ curve must obviously be larger than the BCS curve for all values of $i$.
    Also shown (filled dots) are the exact matrix elements $\bra{N}b_i\ket{N+1}$, 
    calculated using the DMRG approach.
    The inset compares the BCS, the finite-$d$ BCS and the DMRG
    results for a larger grain with $n = 200$ levels. While the finite-$d$ BCS result shows
    excellent agreement with the DMRG result in this regime, the BCS
    result still deviates from the exact result. 
  }
\end{figure}

Next, we discuss the increase of $E_J^0$ in the small-grain limit (ii).
It is due to the fact that the matrix elements
$\bra{N_L}  b_{l} \ket{N_L+1}\bra{N_R+1}  b^\dagger_{r} \ket{N_R}$
that contribute to  $E_J^0$ in \eq{EJ_0} have a different number of electron
pairs in the states acting from the left and on the right.
This fact is neglected in standard BCS theory, where the total number
of pairs is assumed to be macroscopically large anyway. 
When the level spacing $d$ becomes large, however, this
is the main reason for the increase of $E_J^0$:

The increase of $E_J^0$ is easy to understand for the Fermi state ($\lambda = 0$)
and in the BCS limit ($\lambda > 2/\ln N$, see \cite{SchechterDelft01}).
In the Fermi state, the matrix
element $\bra{N}  b_i \ket{N}$ is zero for all values of $i$, but 
$\bra{N}  b_i \ket{N+1}$ gives a contribution of $1$ for the
one level $i = i_N$ that is below the Fermi surface of $\ket{N+1}$ and above
the Fermi surface of $\ket{N}$.
In the BCS case, the matrix element is given by 
$\bra{N}  b_i \ket{N+1} = u^{N}_i v^{N+1}_i$.
The upper indices on $u$ and $v$ indicate the total pair occupation
numbers with respect to which they are taken, with the effect that
$v^{N+1}$ has the chemical potential shifted upwards with respect to $v^N$ by
the level spacing $d$.
Thus, the product  $u^{N}_i v^{N+1}_i$ becomes larger as the level spacing
$d$ increases, as is illustrated in Fig. \ref{uv_fig}.
We shall call this modification of the BCS calculation the ``finite-$d$''
BCS calculation.
\begin{figure}
\epsfig{file=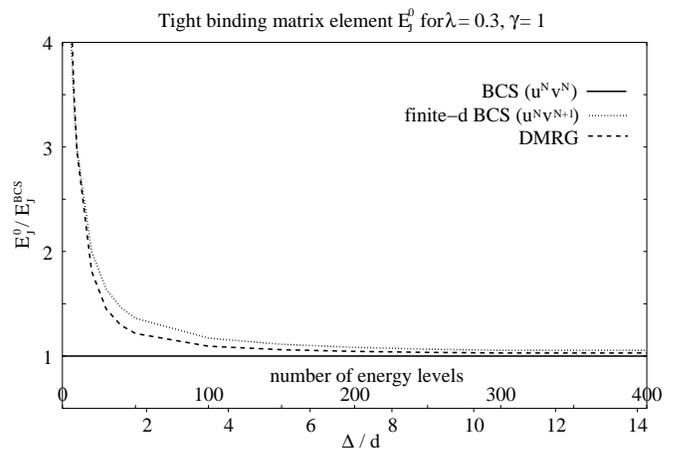,width=\linewidth}
 \caption{\label{EJ_uv_vs_DMRG}
   $E_J^0$ is evaluated in the BCS, the finite-$d$ BCS and the DMRG method. 
   Although BCS theory is not valid in the regime
   of large level spacings $d$, the finite-$d$-BCS
   reproduces, at least qualitatively, the correct behaviour
   seen in the DMRG curve.} 
\end{figure}

In Fig.\ \ref{uv_fig}, the finite-$d$ BCS matrix elements (solid line)
are also compared to the exact values obtained using the DMRG (filled
dots).  The comparison shows that for the levels close to the Fermi
energy (i.e.\ the central level $i_N$ and the next, say, 2 levels),
the finite-$d$ BCS result overestimates the pairing correlations: the
(quasi-)exact DMRG solution is seen to have a more pronounced peak at
the central level $i_N$, whereas the contribution of the neighbouring
two levels is somewhat reduced, resembling, for these levels,
qualitatively more closely the $\lambda =0$ case discussed above.
For the energy levels further away from the Fermi energy than that,
the finite-$d$ BCS calculation is seen to slightly underestimate the
matrix elements.  This is not unexpected, because BCS theory is known
\cite{SchechterDelft01} to underestimate the superconducting
corelations of energy levels much farther away from the Fermi surface
than $\DBCS$, which for the parameters of Fig.\ \ref{uv_fig} is $\DBCS
\approx 0.7$.

$E_J^0$, however, being a weighted sum over all products of these matrix
elements, is nevertheless not so far off in the finite-$d$ BCS
approach even for very small grains, as can be seen in
Fig.\ \ref{EJ_uv_vs_DMRG}. 
This is surprising and somewhat fortituous, since the BCS
theory does not self-consistently describe the grains in the limit
that they are small. 
The reason that the finite-$d$ BCS works so well seems to be that the
underestimation of the matrix elements for level $i_N$ and
and their overestimation for the other levels cancel each other
to a large degree.

In conclusion, the main reason why $E_J^0$ increases as the grains become 
small is very simple: the chemical potential of the
grains shifts due to the finite level spacing whenever a pair tunnels
from one grain to the other.
Note that the BCS ansatz without taking this effect into account is
not accurate near the Fermi energy
even for fairly large grains (see the inset in Fig. \ref{uv_fig}), for which
the finite-$d$ BCS theory agrees perfectly with the DMRG result.

The competition between the finite-size kinetic term on the one hand
and the increase of $E_J^0$ on the other
leads to the reentrant behaviour of $E_J$ as seen in Fig.\
\ref{EJ_tight_fig}. 
This is one of our main results.
Two regimes can be distinguished as a
function of $\DBCS / d$: For very small grains ($\DBCS / d < 1$),
superconducting correlations are only weakly present, but the
1-level effect outlined above leads to a strong enhancement of $E_J^0$
and, therefore, of the Josephson energy $E_J$. 
Despite not being a well-justified approximation in this regime, the
finite-$d$ BCS result nevertheless gives a surprisingly good estimate of the
Josephson energy.
On the other hand, for larger values of $\DBCS / d$ ($> 10$, say),
$E_J^0$ is almost constant and very close to the BCS value.
The kinetic energy term in
\eq{E_kin}, however, reduces the Josephson energy below the BCS
value, and vanishes  only rather slowly.
The reentrant behaviour of $E_J$ occurs at the intermediate region
$1 < \DBCS/d < 10$, in which both effects are competing, and in which
$E_J^0$ is slowly approaching its BCS value from above.

\subsection{Limitations of the tight-binding approach}

The tight-binding approximation, which neglects
pair breaking excitations of the individual grains,
is valid only for small couplings, such that 
$E_J^0$ lies well below the lowest excitation energy $\Delta_\mathrm{sp}$.
In Fig.\ \ref{EJ_tight_fig}, however, $E_J^0$ is seen
to grow strongly with increasing level spacing $d$.
Thus, for sufficiently large $d$, the tight-binding approach invariably becomes
unreliable, and a different method is needed.
In order to complement the tight-binding approach
and to check its quality,
we have thus used the two-grain DMRG solution that does not rely on the
inter-grain coupling being weak.

The DMRG, however, has its own limitations, as was explained in
subsection \ref{twograin_DMRG_subsec}:
Firstly, it requires a pair tunneling Hamiltonian (\ref{H_J_flat}) that
describes a somewhat different model. This implies, of course, that it has to be
compared to a tight-binding model using the same pair tunneling
Hamiltonian as well.
Secondly,
the two-grain DMRG can break down at small couplings if the number
of energy levels is large, for 
precisely the same reason that the tight-binding model works well:
The  correlations between the two grains, which the
DMRG relies on, become very weak.

\begin{figure}
\epsfig{file=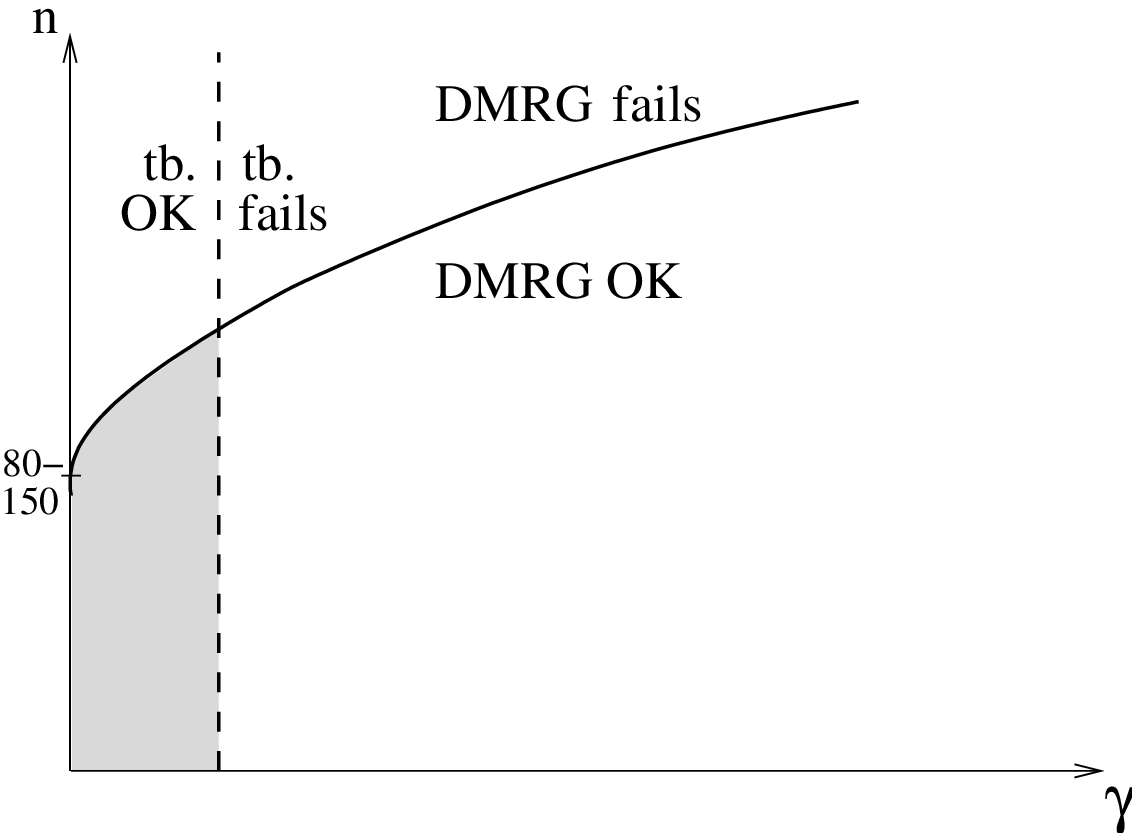,width=.8\linewidth}
 \caption{\label{tb_OK_fig}
   A rough sketch of the regimes of validity for the DMRG  and the
   tight-binding approach in parameter space (inter-grain coupling $\gamma$ vs.\ the
   number $n$ of energy levels).
   There is only a small overlap (shaded region) at small $\gamma$ and
   small $n$, in which both approaches simultaneously work well.}
\end{figure}
\begin{figure}
\epsfig{file=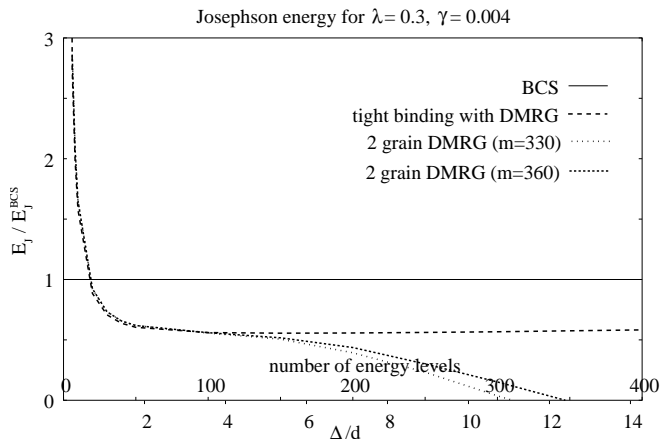,width=\linewidth}
 \caption{ \label{EJ_tb_vs_DMRG_N} The Josephson energy is calculated
   in the tight-binding and in the DMRG approach.
   In agreement with Fig.\ \ref{tb_OK_fig}, both curves agree if the number
   of energy levels $n$ is small, but the DMRG fails for $n > 80 - 100$.} 
\end{figure}
\begin{figure}
\epsfig{file=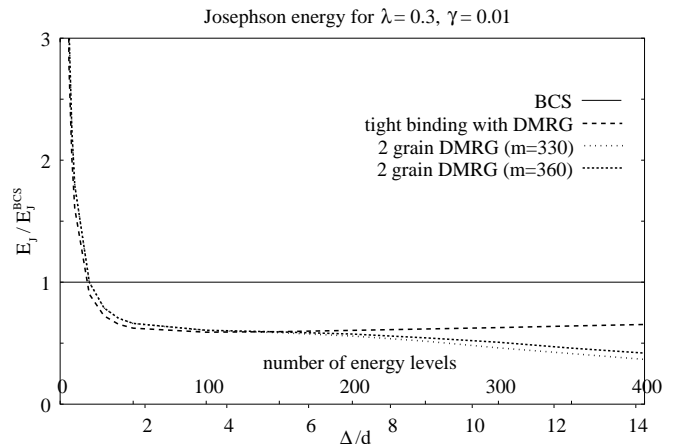,width=\linewidth}
 \caption{\label{EJ_tb_vs_DMRG_N2} Same calculation as in Fig.\ \ref{EJ_tb_vs_DMRG_N}, but at
   larger inter-grain coupling. Now, the DMRG has a somewhat larger range of
   validity.
   For small grains, the DMRG curve lies higher (i.e.\ the
   tight-binding approach is not as good as the DMRG anymore).} 
\end{figure}

The regimes of validity of the two complementary approaches are
schematically depicted in Fig.\ \ref{tb_OK_fig}.
The tight-binding method only works well at small coupling, 
$\Delta_\mathrm{s.p.} \ll E_J$ (region left of dashed line), whereas
the DMRG works well only at large coupling (region right of solid
line).
A simple (analytical) condition for the validity of the DMRG approach cannot be
given, which is why the axes in  Fig.\ \ref{tb_OK_fig} are drawn
without units.
However, the quality of the DMRG approach is found to depend
sensitively on the number of energy levels $n$.
In particular, the DMRG turns out to be reasonably accurate for all values of $\gamma$ down to
0, as long as $n < 80-150$ (depending on other parameters), as is motivated in subsection 
\ref{twograin_DMRG_subsec} and seen in Fig.\ \ref{EJ_tb_vs_DMRG_N} and
Fig.\ \ref{EJ_tb_vs_DMRG_N2}. 

In Fig.\ \ref{EJ_tb_vs_DMRG_N} and Fig.\ \ref{EJ_tb_vs_DMRG_N2},
the tight-binding approximation for the Josephson energy is compared to the 
two-grain DMRG as a function of the grain size, for two different values of the inter-grain coupling
$\gamma$  (corresponding to moving along vertical lines in Fig.\ \ref{tb_OK_fig}).
The Josephson energies are again plotted in units of their BCS
values, now given by \eq{EJ_flat_BCS}.
In Fig.\ \ref{EJ_tb_vs_DMRG_N}, both methods are seen to agree for small numbers
of energy levels, $n < 80 - 100$. 
For larger values of $n$, the two-grain DMRG breaks down, for the
reasons outlined in \ref{twograin_DMRG_subsec}.
The DMRG method itself signals its own breakdown:
Convergence as function of the kept DMRG states $m$ is no longer
achieved, as can already be seen when the two curves shown in Fig.\
\ref{EJ_tb_vs_DMRG_N}, which correspond to $m=330$ and $m=360$, are compared.

Since both the two-grain DMRG and the tight-binding approach are ultimately variational methods, the one that
produces the higher value of $E_J$ (i.e. the lower total condensation
energy) must be the better approximation.
Also in this respect, the DMRG method is seen to be failing for $n >
80-100$ in  Fig.\ \ref{EJ_tb_vs_DMRG_N}, in agreement with Fig.\
\ref{tb_OK_fig}.

Fig.\ \ref{EJ_tb_vs_DMRG_N2} shows the result of a similar calculation
as  Fig.\ \ref{EJ_tb_vs_DMRG_N}, at a higher value of the inter-grain
coupling $\gamma = 0.01$.
Now, the two-grain DMRG is seen to be valid up to somewhat larger values of $n$.
For small $n$,  $n < 100$, the DMRG now produces a higher value of $E_J$,
indicating that in this regime, it produces a better result than the
tight-binding method, as anticipated in  Fig.\ \ref{tb_OK_fig}.

The results from  Fig.\ \ref{EJ_tb_vs_DMRG_N} and Fig.\
\ref{EJ_tb_vs_DMRG_N2} are similar to the ones in Fig.\
\ref{EJ_tight_fig}, where only a tight-binding calculation had been
performed.
In particular, the two-grain DMRG reproduces the increase in $E_J$ for
small values of $n$, corresponding to large level spacing $d$, and
thereby confirms the reentrant behaviour observed in the tight-binding
aproach (cf.\ Fig.\ \ref{EJ_tight_fig}).


\begin{figure} 
\epsfig{file=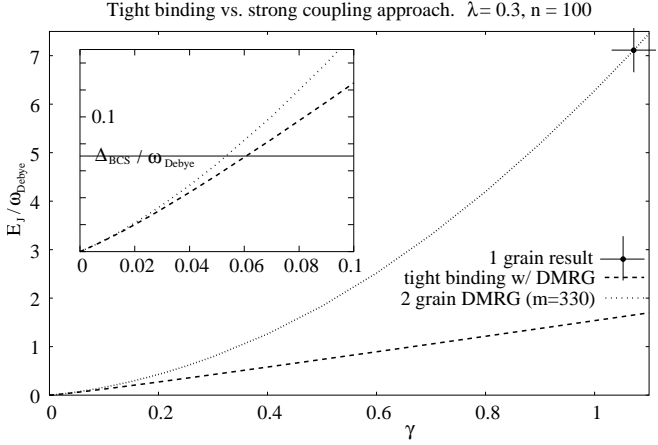,width=\linewidth}\\
\caption{\label{EJ_tb_vs_DMRG_g} The Josephson energy is plotted as a function of
  the inter-grain coupling $\gamma$ in grains that are small enough so that the DMRG approach
  works for all values of $\gamma$ down to zero. Once the coupling is too large
  ($\gamma  \stackrel{>}{\sim} 0.06$), the tight-binding model fails as asserted
  in Fig. \ref{tb_OK_fig}. 
  The inset shows an enlargement for small values of $\gamma$, and
  illustrates the condition $E_J \ll \Dsp \sim \DBCS$ 
  for the tight-binding model to be valid, which was motivated at the
  beginning of section \ref{theory_section}}.
\end{figure}

In Fig.\ \ref{EJ_tb_vs_DMRG_g}, the tight-binding and the two-grain
DMRG results are plotted as a function of the inter-grain coupling $\gamma$  
(corresponding to moving horizontally in Fig.\ \ref{tb_OK_fig}).
The plot is extended to very large values of the inter-grain coupling
$\gamma$ in order to show the point at which the two-grain DMRG can be
compared to the exact result at  $(d / \DBCS) \gamma = \lambda$, which
it reproduces nicely.
We empasize that in the regime of large $\gamma$, some of the physical
assumptions (e.g.~the use of a tunneling Hamiltonian) of our
calculation are not justified anymore, and that the plot in that
regime has no other physical significance than to provide an important
cross check for the DMRG.

The exact result at $(d / \DBCS) \gamma = \lambda$ describes the two
grains as a single superconductor with half the level spacing $d_2 =
d/2$ and with the interaction Hamiltonian 
\be
\label{H2}
H_2 = - \lambda_2 d_2 \sum_{i \in R,L, \; j \in R,L} b^\dagger_i b^\pdag_j,
\ee
and with $\lambda_2 = 2 \lambda$. 
In the large-coupling regime, the Josephson energy 
$E_J = E_{\mathrm{cond},2} - 2 E_{\mathrm{cond},1}$
is entirely dominated by the condensation energy
$E_{\mathrm{cond},2}$ of the large superconductor described by
\eq{H2}, which is much larger than the condensation energies
$2 E_{\mathrm{cond},1}$ of the isolated grains (i.e.~for $\gamma = 0$).
In the BCS limit, 
$E_J \approx E_{\mathrm{cond},2} = \oD n \sinh^2(1/\lambda_2)$.
In particular, $E_J$ is seen to be an extensive quantity, i.e.~ 
$E_J / \oD \propto n$ for the particular choice $\gamma = (\DBCS / d)
\lambda$, for which 
the two superconductors are described as one.
In this case the inter-grain coupling acts like a bulk term (and
no longer as a surface effect), which is manifest in the way that
$\gamma$ scales with 
the system size: $\gamma$ scales no longer as a constant,
but with the volume of the system.

As is evident from  Fig.\ \ref{EJ_tb_vs_DMRG_g},
the tight-binding method, which is only applicable at very small values of
$\gamma$, ceases to be valid long before the point $(d / \DBCS) \gamma
= \lambda$ is reached.
The inset of Fig.\ \ref{EJ_tb_vs_DMRG_g} shows an enlargement of the
main figure for small $\gamma$.  
It is seen that for $E_J^0 \ll \DBCS$, the results from the tight-binding 
method and from the DMRG agree, as expected. 
\\
{\small
This work was supported by the DFG Program ``Semiconductor and Metallic 
Clusters''.
US and DG acknowledge support through a Gerhard Hess prize of the Deutsche
Forschungsgemeinschaft.
We thank 
G.~Falci, Y.~Imry, S.~Kleff, C.~Kollath, M.~Schechter, and G.~Sierra
for discussions.
}

\begin{appendix}
\section{\label{DMRG_app} The DMRG algorithm in energy space}

In this appendix, some technical aspects of the DMRG procedure for
approximating the ground state $\ket{\psi}$ are explained. 
There are excellent pedagogical reviews of the DMRG algorithm to be
found\cite{WhiteNoack99,Peschel99},
as well as a description of its application to
superconducting grains\cite{SierraDukelsky00,DukelskySierra99}, so
we only highlight  the key concepts of the DMRG
algorithm for reducing the Hilbert space.
Then, the full DMRG algorithm as applied in energy space is sketched.
Finally, a few peculiarities are mentioned that are of relevance
when the algorithm is applied to the problem of two coupled superconductors.

\begin{figure}
\epsfig{file=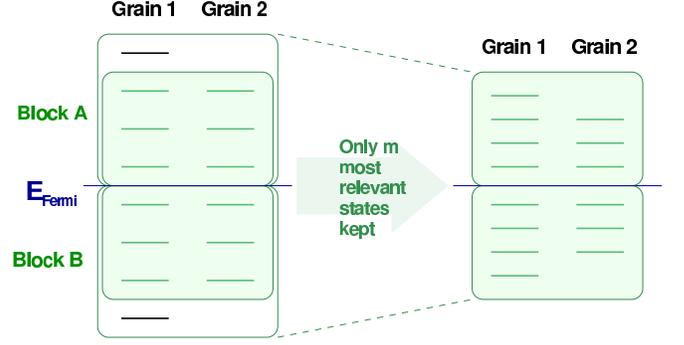,width=\linewidth}
 \caption{\label{densmat_fig}
   Sketch of the procedure for projecting out the relevant
   states in the case of the two-grain DMRG. 
   The shading indicates the part of the Hilbert space where
   only a limited number of states are kept.
   First, a new level is added on grain 1 (left part of
   figure). Then, the $m$ most relevant states are projected out and
   kept (right part). Then, a new level on grain 2 is added (not shown).}
\end{figure}

First, we will give an account of the procedure that projects out a
reduced number of basis states.
The Hilbert space is divided into two blocks $A$ of states below and
$B$ of states above the Fermi surface, as depicted
in Fig.\ \ref{densmat_fig}, each being represented by the respective
basis state $\ket{i}_A$ and $\ket{j}_B$.
A general many-body state is expressed as 
$\sum_{ij} \psi_{ij} \ket{i}_A \otimes \ket{j}_B$.
The goal is to find a reduced number $m$ of most relevant states 
$\ket{u_\alpha}_A$ and $\ket{u_\beta}_B$,
in the sense that they allow
for the best approximation of the state $\ket{\psi}$, such that the
norm
$\left| 
\ket{\psi} - 
\sum_{i j} \psi_{\alpha \beta} \ket{u_\alpha}_A \otimes
\ket{u_\beta}_B
\right|$
is minimized,
when variation over both $ \psi_{\alpha \beta}$ and the states
$\ket{u_\alpha}_A, \; \ket{u_\beta}_B$ are allowed, but only $m$
states per block are to be kept.
It turns out that the states with this property are precisely those
eigenstates of the reduced density matrix of the respective block ($A$ or
$B$) that correspond to the $m$ largest eigenvalues\cite{WhiteNoack99}.
Of course, the larger $m$ is, the more accurate the algorithm becomes,
until convergence is achieved.
Typical values for $m$ are $m \sim 100-400$.

The prescription for the DMRG algorithm is the following:
(i) Start with only a few (2 or 3, say) energy levels, few enough that
the exact basis of the many-body system can be kept explicitly.
(ii) Add an additional energy level to block $A$ and $B$, as depicted in
Fig.\ \ref{densmat_fig} for the case of the two-grain DMRG.
Construct a basis $\ket{u_\alpha}_A$ for block $A$, using the basis states from
the previous step and
the exact basis of the newly added energy level. 
Do the same with block $B$.
(iii) Calculate the target state $\ket{\psi}$, in our case the ground state
of the BCS Hamiltonian, within the present Hilbert space.
(iv) Calculate the reduced density matrix of $\ket{\psi}$ for
block $A$ and $B$, say $\rho_A$ and $\rho_B$,
by tracing out the full density matrix 
$\ket{\psi} \bra{\psi}$ over the respective other block. 
Find the  $m$ eigenvectors $\ket{u_\alpha}_A$, $\ket{u_\beta}_B$, $\alpha, \beta = 1..m$,
corresponding to the $m$ largest eigenvalues of  $\rho_A$ and $\rho_B$. 
Those are the states to be kept as basis states.
(v) Transform all operators to the new basis.
If the blocks $A$ and $B$ are related by a symmetry, it may be sufficient
to calculate only one set of states $\ket{u_\alpha}$.
Continue with step (ii) and iterate, until the final number of energy levels is reached.

In step (iii), the ground state $\ket{\psi}$ is found using the Lanczos procedure,
which is very efficient due to the sparse nature of the Hamiltonian,
but which requires many multiplications of a state with the Hamiltonian.
Since the Hamiltonian is a sparse but extremely large matrix 
(of order $m^2 \times m^2$), it is essential not to store it as a whole, but to reconstruct
it from simple operators acting only on the blocks $A$ and $B$
when the multiplication is performed.
For this to be numerically possible, it is
necessary that the interactions between the blocks factorize 
to a large degree, such that they  can be expressed as a sum of only a
few terms.
In real-space DMRG, this is
always the case as long as the interactions are more or less local,
but the long-range interactions in energy space do
not always factorize.
Luckily, the reduced BCS interaction does factorize nicely:
$H_{BCS} = 
- \lambda (B_A^\dagger B_A^\pdag + B_A^\dagger B_B^\pdag + A \leftrightarrow B)$,
where $B_{A,B} = \sum_{i \in A,B} c_{i \uparrow} c_{i \downarrow}$.
A similar factorization is possible for the inter-grain coupling (\ref{H_J_flat}) in the
two-grain DMRG, but not for (\ref{H_J}).

It is also essential for numerical efficiency to make use of conserved
quantum numbers.
In our case, due to particle number conservation, it is not necessary
to keep all the $m^2$ states $\ket{u_\alpha}_A \otimes \ket{u_\beta}_B$ as a basis.
In our algorithm, we keep track of the number $l_{\alpha}, l_\beta$ of particle or hole
excitations associated with each basis vector $\ket{u_\alpha}_A, \ket{u_\beta}_B$,
respectively.
Then, only the states $\ket{u_\alpha}_A \otimes \ket{u_\beta}_B$ have to be kept
for which 
\be
\label{delta_l}
l_\alpha - l_\beta = l_\mathrm{tot}, 
\ee
where $l_\mathrm{tot}$ is
the deviation of the total electron pair number from half filling.

In the tight-binding calculation, taking the matrix element 
$\bra{n} b_i \ket{n+1}$ involves approximating two states
simultaneously, namely the ground states $\ket{n}$ and $\ket{n+1}$
that correspond to the respective number of electron pairs $n$ and
$n+1$.
This is simply done by calculating both states in step (iii), and by
taking the reduced density matrix of the mixed state with equal weight
in step (iv).

In the two-grain DMRG, the calculations are performed in the regime
that two states, $\ket{\nu}$ and $\ket{\nu+1}$, as defined in
\eq{nu_state}, are degenerate.  This is done by setting the offset
between the energy levels on the left and the right grain to zero, and
by including one more electron pair than there would be at half
filling, which amounts to setting $l_\mathrm{tot} = 1$ in
\eq{delta_l}.  This extra pair can, then, be on the left or the right
grain at equal energy cost.

One complication arises away from half filling (i.e. when
$l_\mathrm{tot} \neq 0$): When, in step (iv), the reduced basis of one
block (block $A$, say) is calculated by tracing over the states in the
other block, the part of the trace relevant for the states with
quantum number $l_\alpha$ is, due to \eq{delta_l}, performed over
states which carry a different quantum number $l_\beta$.  The
dimensionality of the two subspaces ${\cal{H}}(l_\alpha)$ and
${\cal{H}}(l_\beta)$ spanned by the part of the reduced density matrix
with the respective quantum numbers might be quite different.
However, the rank of the reduced density matrix used in step (iv) is
limited by the dimension of the space over which the trace is
performed, and therefore, the DMRG only works well as long as the
dimension of ${\cal{H}}(l_\beta)$ is larger than the number of states
with quantum number $l_\alpha$ to be kept.  This is not guaranteed
away from half filling, i.e.\ when $l_\alpha \neq l_\beta$.

The problem is solved by mixing a small
part (20\%) into the reduced density matrix that corresponds to the
ground state at half filling ($l_\mathrm{tot} = 0$). 
This state will have a similar information content as the target state away
from half filling, as far as the
relevant basis vectors are concerned, and adds enough
to the rank of the reduced density matrix for the DMRG to work well.

In the two-grain DMRG,
the energy levels are added one by one as depicted in Fig.\ 
\ref{densmat_fig}:
First levels on grain 1, and only afterwards levels
on grain 2 are added.
They are added one by one in order to keep the Hilbert space as small
as possible.
It is also possible and, in fact, would be more symmetric, 
to add both levels at once, but only at the
cost of having the Hilbert space larger by a factor of 4.
As it turns out, it is numerically more efficient (yielding higher
accuracy at the same computation time) to add the levels one by one. 

\end{appendix}


\begin{thebibliography}{10}

\bibitem{AmbegaokarBaratoff63}
V.~Ambegaokar and A.~Baratoff.
\newblock Tunneling between superconductors.
\newblock {\em Phys. Rev. Lett.}, 10:486, 1963.

\bibitem{Anderson59}
P.~W. Anderson.
\newblock {\em J. Phys. Chem. Solids}, 11:28, 1959.

\bibitem{LikharevAverin91}
D.~Averin and K.~Likharev.
\newblock Single electronics: A correlated transfer of single electrons and
  cooper pairs in systems of small tunnel junctions.
\newblock In B.~Altshuler, P.~Lee, and R.~Webb, editors, {\em Mesoscopic
  phenomena in solids}. Elsevier, 1999.

\bibitem{BlackTinkham96}
C.~Black, D.~Ralph, and M.~Tinkham.
\newblock Spectroscopy of the superconducting gap in individual nanometer-scale
  aluminum particles.
\newblock {\em Phys. Rev. Lett.}, 76:688, 1996.

\bibitem{deGennes99}
P.-G. de~Gennes.
\newblock {\em Superconductivity of metals and alloys}.
\newblock Perseus books, 1999.

\bibitem{DukelskySierra99}
J.~Dukelsky and G.~Sierra.
\newblock Density matrix renormalization group study of ultrasmall
  superconducting grains.
\newblock {\em Phys. Rev. Lett.}, 83:172, 1999.

\bibitem{Ferrell88}
R.~Ferrell.
\newblock Josephson effect and anderson's theorem.
\newblock {\em Phys. Rev. B}, 38:4984, 1988.

\bibitem{HofstetterLukin02}
W.~Hofstetter, J.~Cirac, P.~Zoller, E.~Demler, and M.~Lukin.
\newblock High-temperature superfluidity of fermionic atoms in optical
  lattices.
\newblock {\em Phys. Rev. Lett.}, 89:220407, 2002.

\bibitem{IansitiLobb89}
M.~Iansiti, M.~Tinkham, A.~Johnson, W.~Smith, and C.~Lobb.
\newblock Charging effects and quantum properties of small superconducting
  tunnel junctions.
\newblock {\em Phys. Rev. B}, 39:6465, 1989.

\bibitem{KnorrLeslie73}
K.~Knorr and J.~Leslie.
\newblock Metal-insulator-metal tunnel junctions.
\newblock {\em Solid State Commun.}, 12:615, 1973.

\bibitem{LinksHibberd02}
J.~Links and K.~Hibberd.
\newblock Integrable coupling in a model for josephson tunneling between
  non-identical bcs systems.
\newblock {\em cond-mat/0206331}, 2002.

\bibitem{MatveevShekhter93}
K.~Matveev, M.~Gisself{\"a}lt, L.~Glazman, M.~Jonson, and R.~Shekhter.
\newblock Parity-induced suppression of the coulomb blockade of josephson
  tunneling.
\newblock {\em Phys. Rev. Lett.}, 70:2940, 1993.

\bibitem{NakamuraTsai99}
Y.~Nakamura, Y.~Pashkin, and J.~Tsai.
\newblock Coherent control of macroscopic quantum states in a
  single-cooper-pair box.
\newblock {\em Nature}, 398:786, 1999.

\bibitem{Peschel99}
I.~Peschel.
\newblock {\em Density matrix renormalization : a new numerical method in
  physics}.
\newblock Springer, 1999.

\bibitem{RichardsonSherman64}
R.~Richardson and N.~Sherman.
\newblock {\em Nucl. Phys.}, 52:221, 1964.

\bibitem{SchechterDelft01}
M.~Schechter, Y.~Imry, Y.~Levinson, and J.~von Delft.
\newblock Thermodynamic properties of a small superconducting grain.
\newblock {\em Phys. Rev. B}, 63:214518, 2001.

\bibitem{SierraDukelsky00}
G.~Sierra and J.~Dukelsky.
\newblock Crossover from bulk to few-electron limit in ultrasmall metalic
  grains.
\newblock {\em Phys. Rev. B}, 61:12302, 2000.

\bibitem{Tinkham96}
M.~Tinkham.
\newblock {\em Introduction to superconductivity}.
\newblock McGraw-Hill, 1996.

\bibitem{DelftRalph01}
J.~von Delft and D.C. Ralph.
\newblock Spectroscopy of discrete energy levels in ultrasmall metallic grains.
\newblock {\em Physics Reports}, 345:61, 2001.

\bibitem{WhiteNoack99}
S.R. White and R.M. Noack.
\newblock Density matrix renormalization group.
\newblock In I.~Peschel, editor, {\em Density matrix renormalization :a new
  numerical method in physics}. Springer, 1999.

\bibitem{ref:timereversed}
  In general, $b_i$ annihilates a pair of electrons $c_{i \uparrow}
  c_{\bar i \downarrow}$ in time-reversed states $|i, \uparrow
  \rangle$ and $|\bar i, \downarrow \rangle$; for the present context
  of nanograins in the absence of a magnetic field, however, we may
  take $i = \bar i $.

\end{thebibliography}

\end{document}